\documentclass[twocolumn,showpacs,superscriptaddress,prb]{revtex4}

\usepackage{graphicx}
\usepackage{dcolumn}
\usepackage{bm}
\usepackage{amsmath}
\usepackage{amsfonts}
\usepackage{amssymb}

\begin{document}

\title{Two-Mode Squeezed States and Entangled States of Two Mechanical Resonators}
\author{Fei Xue}
\affiliation{CREST, Japan Science and Technology Agency (JST),
Kawaguchi, Saitama 332-0012, Japan}
\affiliation{Frontier Research
System, The Institute of Physical and Chemical Research (RIKEN),
Wako-shi, Saitama 351-0198, Japan}
\author{Yu-xi Liu}
\affiliation{CREST, Japan Science and Technology Agency (JST),
Kawaguchi, Saitama 332-0012, Japan}
\affiliation{Frontier Research
System, The Institute of Physical and Chemical Research (RIKEN),
Wako-shi, Saitama 351-0198, Japan}
\author{C. P. Sun}
\affiliation{Institute of Theoretical Physics, Chinese Academy of
Sciences, Beijing, 100080,  China}
\author{Franco Nori}
\affiliation{CREST, Japan Science and Technology Agency (JST),
Kawaguchi, Saitama 332-0012, Japan}
\affiliation{Frontier Research
System, The Institute of Physical and Chemical Research (RIKEN),
Wako-shi, Saitama 351-0198, Japan}
\affiliation{Center for
Theoretical Physics, Physics Department, Center for the Study of
Complex Systems, The University of Michigan, Ann Arbor, Michigan
48109-1040, USA}


\date{\today }

\begin{abstract}
We study a device consisting of a dc-SQUID with two sections of its
loop acting as two mechanical resonators. An analog of the
parametric down-conversion process in quantum optics can be realized
with this device. We show that a two-mode squeezed state can be
generated for two overdamped mechanical resonators, where the
damping constants of the two mechanical resonators are larger than
the coupling strengths between the dc-SQUID and the two mechanical
resonators. Thus we show that entangled states of these two
mechanical resonators can be generated.
\end{abstract}

\maketitle

\section{Introduction}

\label{sec:introduction}

Motivated by their relevance to quantum information, coherent
quantum behavior of macroscopic solid-state devices are of great
interest. Quantized energy levels, coherent time evolution,
superposition and entangled states---have all been observed in
various solid-state devices, such as quantum dots and
superconducting quantum interference devices (SQUIDs).
Nanomechanical resonators
(NAMRs)~\cite{Clelandbook2002,Blick2002,Blencowe2004,Schwab2005PT}
with frequencies as high as Giga Hertzs can now be
fabricated~\cite{Huang2003,Gaidarzhy2005,Schwab2005comment,Gaidarzhy2005reply}.
At milli-Kelvin temperatures, such mechanical resonators are
expected to exhibit coherent quantum behavior. In order to detect
and control mechanical resonators, some transducer methods must be
used. These include optical methods, magnetomotive techniques, and
couplings to single electron
transistors)~\cite{Clelandbook2002,Blick2002,Blencowe2004,Schwab2005PT}.
A novel design of mechanical qubits based on buckling nanobars was
recently studied in Ref.~\onlinecite{Savel'ev2006NJP}. Also, buckled
modes analogous to buckled-bars have been proposed for magnetic
nanostructures~\cite{Savel'ev2004Dec}, and mechanical bars bent by
electric fields have been considered, e.g.,
Ref.~\onlinecite{Nishiguchi2003Sep}. Moreover, these quantum
mechanical nanobars can exhibit behavior similar to superconducting
quantum circuits~\cite{Savel'ev2007}.

The quantization of NAMRs has been studied by coupling NAMRs to a
superconducting charge
qubits~\cite{Armour2002,Wang2004,Zhang2005,Sun2006,Wei2006}. By
controlling the charge qubit, a NAMR can be prepared into different
quantum states. Also, charge qubits can also be used to measure the
quantum states of the NAMRs.  Quantum nondemolition measurements of
a NAMR were studied with an rf-SQUID acting as a transducer between
the NAMR and an LC resonator~\cite{Buks2006oct}. It was shown that a
strong coupling cavity QED regime can be realized for a NAMR and a
superconducting flux qubit~\cite{Xue2007NJP} or a NAMR with magnetic
tip coupled to an electron spin~\cite{Xue2007prb}.

For their usages of sensitive displacement detection beyond the
standard quantum limit, single-mode squeezed states of a
nanomechanical resonator were theoretically studied. It was shown
that squeezed states of the nanomechanical resonator can be
generated by either periodically flipping a superconducting charge
qubit coupled to it~\cite{Zhou2006Dec} or by measuring the
superconducting charge qubit coupled to it~\cite{Rabl2004Nov}. The
squeezing of the nanoresonator state can also be produced by
periodically measuring its position by a single-electron
transistor~\cite{Ruskov2005Nov}.

Two-mode squeezed states, when these two modes are from two
spatially-separated macroscopic objects, are macroscopic entangled
states. The generation of these entangled states of macroscopic
objects are of fundamental interest. Several protocols have been
proposed to entangle two tiny mirrors with the assistance of
photons~\cite{Mancini2002,Pirandola2006}. Here we study a device
consisting of a dc-SQUID with two opposite sections of the SQUID
loop suspended from the substrate. The suspended parts, shaped as
doubly-clamped beams, can be approximated as NAMRs. The magnetic
flux threading the loop of the dc-SQUID is modulated by the
displacements of both NAMRs. Then the dynamics of the dc-SQUID is
modified by the NAMRs. We study how the potential energy of the
dc-SQUID is modified by the displacement of the NAMRs. We show that
the nonlinear coupling between a dc-SQUID and the NAMRs, where the
dc-SQUID is approximated as a quantum harmonic oscillator, offers a
flexible method for the detection and control of NAMRs.
Specifically, we discuss two-mode squeezed states of these two NAMRs
through an analog of the \textit{two-mode parametric
down-conversion} process in quantum optics. We show here that
two-mode squeezed states of the two NAMRs can be obtained even when
the couplings between the dc-SQUID and the two NAMRs are weaker than
their damping rates. In contrast to this, for the proposal studied
in Ref.~\onlinecite{Eisert2004} the entanglement between the NAMRs
decreases rapidly with time. Other forms of squeezing of mechanical
oscillators (phonons) have been studied about a decade ago in
Refs.~\onlinecite{Hu1995,Hu1996,Hu1997,Hu1999}.

This paper is organized as follows. At the beginning of
Sec.~\ref{sec:the device} our device is described. Then, after we
study the potential energy of the dc-SQUID, the Hamiltonian of the
device is presented. The interaction Hamiltonian between a dc-SQUID
and two NAMRs is complicated and has many terms. However, if the
frequency of the dc-SQUID is properly chosen by the bias current of
the dc-SQUID, then only a few terms dominate the dynamics of the
coupled system, which is illustrated by writing the interaction
Hamiltonian in the interaction picture. Then, in
Sec.~\ref{sec:squeezing} we study a special case where, under an
appropriate choice of the parameters, the interaction Hamiltonian is
simplified to study the two-mode parametric down-conversion process
in the device. Squeezed states of the two NAMRs are studied by the
Heisenberg-Langevin method. Conclusions are given in
Sec.~\ref{sec:discussions}.

\section{Coupling a dc-SQUID with two nanomechanical resonators}

\label{sec:the device}

\begin{figure}[tp]
\centering
\includegraphics[bb=200 360 430 525,width=6cm,clip]{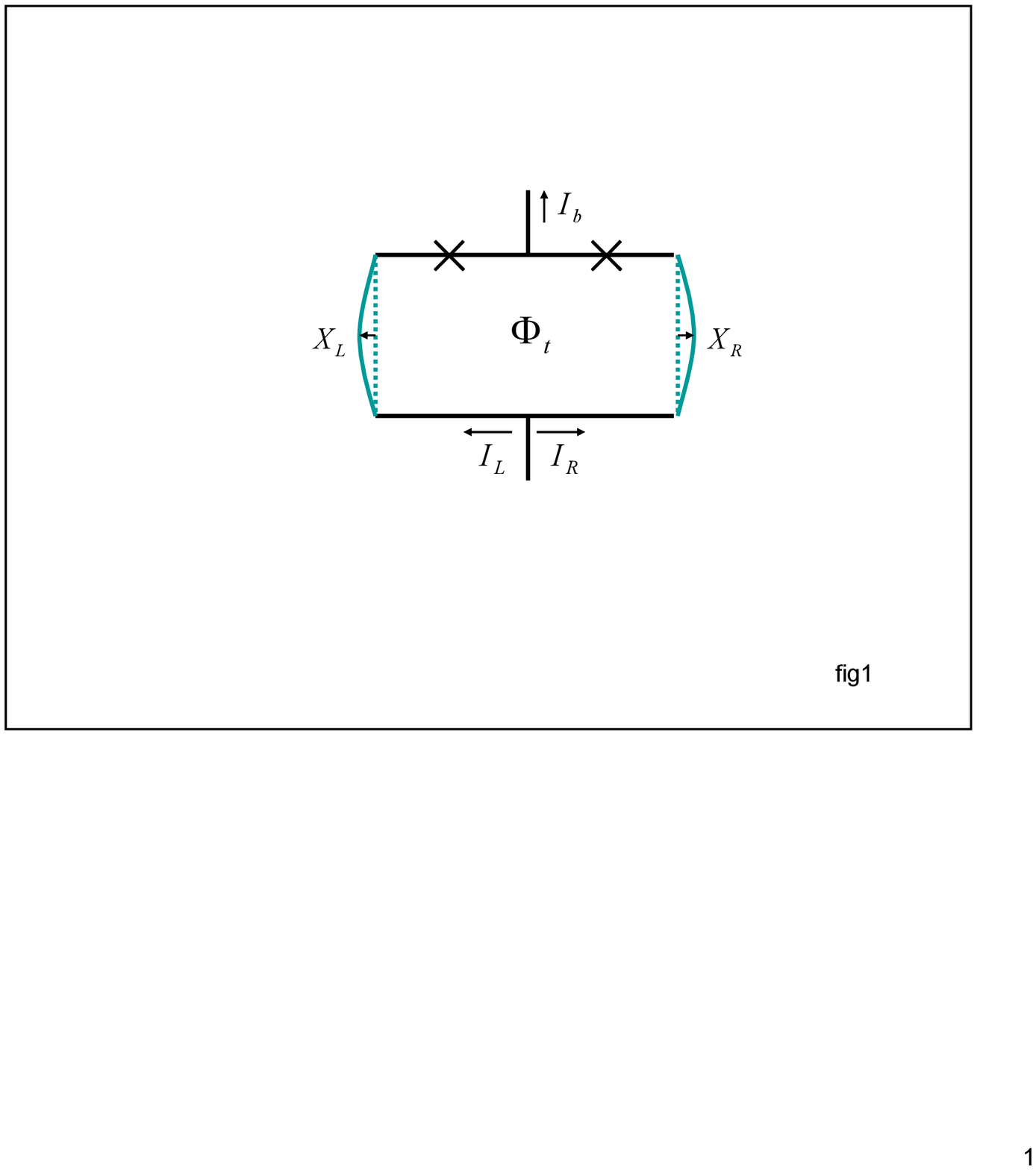}
\caption{(Color online) Schematic diagram of the top view of the
device considered here. It consists of a rectangular-shaped dc-SQUID
and two mechanical resonators shown in blue on the left and right
sides. These two opposite segments of the dc-SQUID are freely
suspended and are treated as two nanomechanical resonators (NAMRs).
The dotted lines indicate the equilibrium positions of the left and
right NAMRs. Each \textquotedblleft $\times$\textquotedblright in
the top segment of the loop represents a Josephson junction.
$X_L$~($X_R$) is the displacement of the center of the left (right)
resonator. The two NAMRs could be located sufficiently ``far apart''
by using a SQUID with an appropriate aspect ratio. This could
provide EPR-type correlations on the two NAMRs located sufficiently
``far apart'' for wide-enough SQUIDs. \label{fig:device} }
\end{figure}

The device we studied is schematically illustrated in
Fig.~\ref{fig:device}. It consists of a dc-current-biased SQUID with
rectangular shape and with two mechanical resonators. The left and
right sides of the SQUID are suspended from the substrate and form
the two mechanical oscillators, our NAMRs. We assume here that these
two doubly-clamped beams vibrate in their fundamental flexural modes
and in the plane of the SQUID loop. We use the following notations
$I_{L}$ ($I_{R}$) for the current in the left (right) Josephson
junction, and $\varphi _{L}$ ($\varphi _{R}$) for the phase drop in
the left (right) Josephson junction. The two Josephson junctions are
assumed to be identical and have the same critical current $I_{c}$.
Thus the bias current $I_{b}$ of the dc-SQUID has the form
\begin{equation}
I_{b}=I_{c}\left( \sin \varphi _{L}+\sin \varphi _{R}\right)
\,\,\text{.} \label{eq:bias current1}
\end{equation}%
We assume that the inductance of the dc-SQUID loop is negligibly
small, and thus the magnetic energy of the circulating current in
the dc-SQUID loop is neglected. Thus the voltage drop over the two
junctions is zero. Therefore, $\varphi _{R}-\varphi _{L}=\varphi
_{t}$, where $\varphi _{t}$ is the phase related to total magnetic
flux $\Phi _{t}$ threading the dc-SQUID loop
\begin{equation}
\varphi _{t}=2\pi \frac{\Phi _{t}}{\Phi _{0}}\,\,\text{.}
\end{equation}%
Here $\Phi _{0}=h/2e\ $is flux quantum. Introducing two new variables
\begin{subequations}
\begin{eqnarray}
\varphi &=&\frac{1}{2}\left( \varphi _{R}+\varphi _{L}\right) \,\,\text{,} \\
\varphi _{-} &=&\frac{1}{2}\left( \varphi _{R}-\varphi _{L}\right)
\,\,\text{,}
\end{eqnarray}%
\end{subequations}
and taking into account that $\varphi _{-}=\varphi _{t}/2$, the bias
current in Eq.~(\ref{eq:bias current1}) can be written as
\begin{equation}
I_{b}=2I_{c}\sin \varphi \cos \frac{\varphi _{t}}{2}\,\,\text{.}
\label{eq:bias current2}
\end{equation}%
It is here assumed that $X_{L}$ ($X_{R}$) is the amplitude for the
fundamental flexural mode of the left (right) beam. Let $B_{L}$
($B_{R}$) be the magnetic field normal to the plane of the SQUID
loop near the left (right) mechanical beam and $\Phi _{b}$ the
external applied magnetic flux threading perpendicularly the
dc-SQUID loop when $X_{L}=X_{R}=0$. It is assumed that $B_{L}$
($B_{R}$) is constant in the oscillating region of the left (right)
beam. Then, the total magnetic flux threading the dc-SQUID loop is
given by
\begin{equation}
\Phi _{t}=\Phi _{b}+\Phi _{X}\,\,\text{,}
\end{equation}%
where $\Phi _{X}$ is the additional magnetic flux when the two NAMRs
are displaced from their equilibrium positions:
\begin{equation}
\Phi _{X}=B_{L}\,X_{L}\,l+B_{R}\,X_{R}\,l\,\,\text{.}
\label{eq:flux-NAMRs}
\end{equation}%
Here, $l$ is the effective length of the left and right beam. $l$ is
defined as $l\equiv S_L/X_L$, where $S_L$ is the area between the
equilibrium position of the NAMR and its bent configuration. Namely
the area $S_L$ spans the region between the blue dashed line and the
blue bent line in Fig.~\ref{fig:device}.
Equation~(\ref{eq:flux-NAMRs}) indicates that the variables of the
two NAMRs enter in the dynamics of the dc-SQUID by influencing the
flux threading the dc-SQUID. The influence of the two NAMRs on the
dynamics of the SQUID can also be revealed quantitatively in the
potential energy of the dc-SQUID, since this is also a function of
the displacements of the two NAMRs. Thus, we first study the
potential energy of the SQUID and afterwards the entire Hamiltonian
of the coupled system.

\subsection{Potential Energy of the Vibrating dc-SQUID}

\begin{figure}[tp]
\centering
\includegraphics[bb=130 260 390 570, clip, width=8cm]{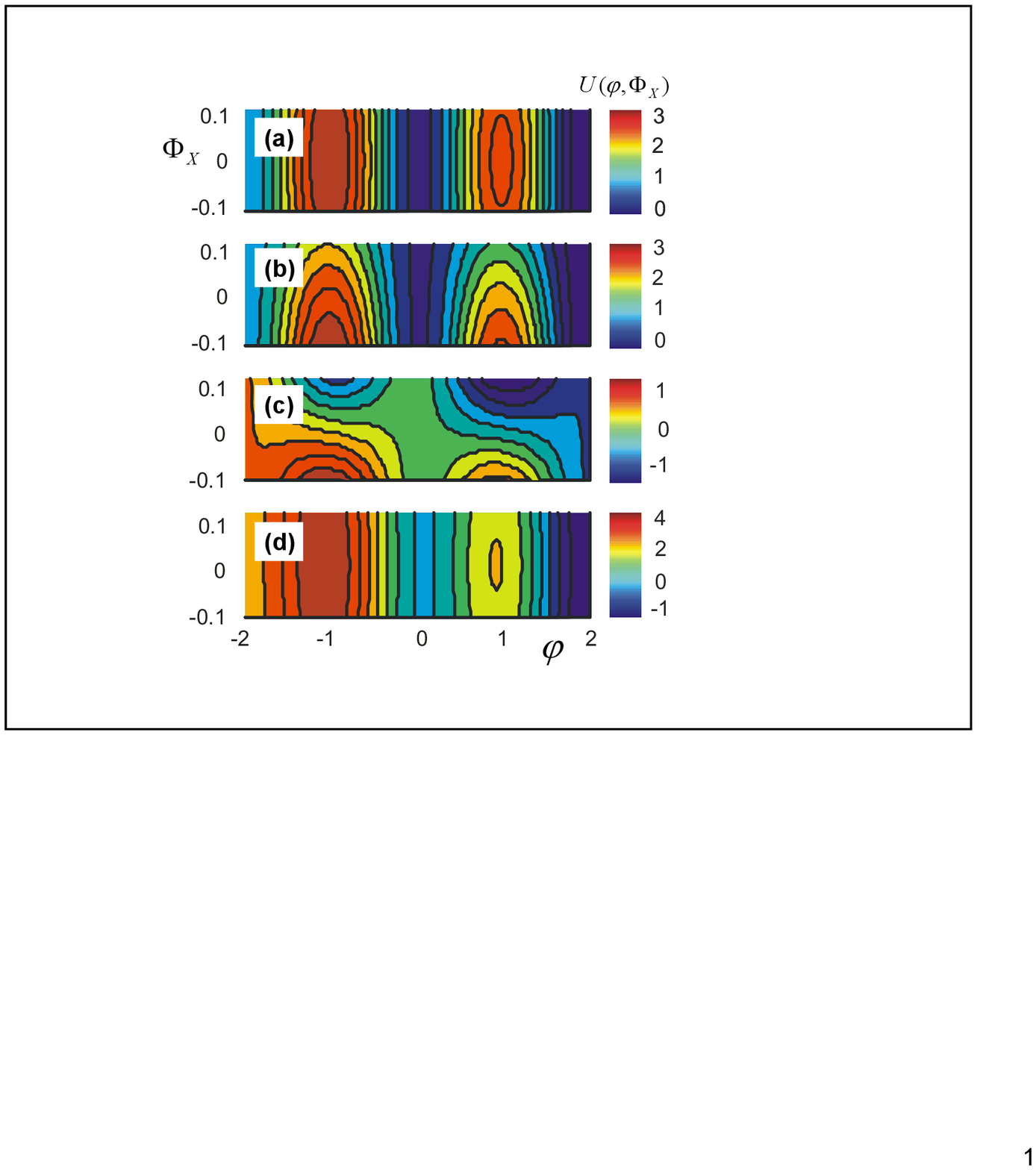}
\caption{(Color online) The potential energy $U(\varphi ,\Phi _{X})$
(scaled by $E_J$) of the dc-SQUID as a function of the phase
variable $\varphi$ and the magnetic flux $\Phi _{X}$ originating
from the displacements of the two NAMRs. The red color represents
higher potential energy $U$, and the blue color represents lower
potential energy. Both $\varphi$ and $\Phi _{X}$ are shown in units
of $\pi/\Phi _{0}$. In (a)-(c), the bias magnetic flux $\Phi _{b}$
threading the loop of the dc-SQUID is set at $2n\Phi _{0}$, $(
2n+\frac{1}{4}) \Phi _{0} $, and $( 2n+\frac{1}{2}) \Phi _{0}$,
respectively; and the bias currents are all set at $I_b=0.1 I_c$. In
(d), $\Phi _{b}=2n\Phi _{0}$ and $I_b=0.5 I_c$.
\label{fig:potential}}
\end{figure}

The potential energy of the dc-SQUID is
\begin{eqnarray}
&&U\left( \varphi ,\Phi _{X}\right)  \notag \\
&=&-2E_{J}\cos \left( \pi \frac{\Phi _{b}}{\Phi _{0}}+\pi \frac{\Phi _{X}}{%
\Phi _{0}}\right) \cos \varphi  -\frac{I_{b}}{I_{c}}%
E_{J}\,\varphi \,\,\text{,} \label{eq:potential energy}
\end{eqnarray}%
where $E_{J}=\hbar I_{c}/(2e)$ is the Josephson energy of the junction~\cite%
{You2005PT,Wendin2005}. To have an idea of the situation under which
the dc-SQUID can be described by a quasi-particle in a quadratic
potential, in Fig.~\ref{fig:potential}~(a)-(d) we plot this
potential energy (\ref{eq:potential energy}) for various values of
the bias magnetic flux $\Phi _{b}$ and the bias current $I_b$. Since
$\Phi _{X}/\Phi _{0}\ll 1$ in the case of experiments using a GHz
NAMR, here we focus on the limit $\Phi _{X}/\Phi _{0}<0.1$ in
Fig.~\ref{fig:potential}. A particle in a quadratic potential can be
described by a harmonic oscillator when its kinetic energy is much
smaller than the barrier of the potential. We notice that, for the
bias magnetic flux $\Phi _{b}=2n\Phi _{0}$, with $n$ being an
integer, the phase variable $\varphi$ falls in a potential well when
the NAMRs oscillate around their equilibrium points. It is possible
to approximate the dynamics of $\varphi$ as a harmonic oscillator.
The charging energy of the dc-SQUID, $E_{c}\equiv\left( 2e\right)
^{2}/\left( 2C_{J}\right) $, is assumed here to be much smaller than
the modified Josephson energy $(\cos q_{0})E_{J}$ of the dc-SQUID.
Here, $q_{0}=\sin ^{-1}\left(I_{b}/2I_{c}\right)$ is a value of
$\varphi$ which corresponds to one of the minima of the potential
energy $U(\varphi, \Phi_{X})$, when the two NAMRs are at their
equilibrium positions. $C_{J}$ is the capacitance of the left and
right Josephson junctions.

We then expand the potential $U\left( \varphi ,\Phi _{X}\right) $
near one of its minimum points $(\varphi ,\Phi _{X})=(q_{0},0)$.
When $\Phi _{X}/\Phi _{0}\ll 1$, and $\Phi _{b}=2n\Phi _{0}$, the
first cosine in Eq.~(\ref{eq:potential energy}) depends weakly on
$\Phi_{X}$. Up to second order in $\Phi _{X}$, we can also expand
After shifting the origin of $\varphi $ to $q_{0}$, and omitting the
constant terms, the potential energy in Eq.~(\ref{eq:potential
energy}) becomes
\begin{eqnarray}
&&U\left( \varphi ,\Phi _{X}\right)  \notag \\
&=&E_{J}(\cos q_{0})\,\varphi ^{2}-E_{J}\left( 1-\cos q_{0}\right)
\left( \pi
\frac{\Phi _{X}}{\Phi _{0}}\right) ^{2}  \notag \\
&&-E_{J}\left[ (\sin q_{0})\varphi +\frac{1}{2}(\cos q_{0})\varphi
^{2}\right] \left( \pi \frac{\Phi _{X}}{\Phi _{0}}\right)
^{2}\,\,\text{.} \label{eq:potential}
\end{eqnarray}%
Therefore, if the first term in the above potential is much larger
than the other two terms, the dynamics of $\varphi $ will still be
well described by a harmonic oscillator.

The higher order terms, such as $\pi^4 \Phi _{X}^4/ \left( 24 \Phi
_{0} ^4 \right) $, in the expansion of $\cos\left( \pi {\Phi
_{X}}/{\Phi _{0}}\right)$ are negligibly small for the situation
considered in our paper, when ${\Phi _{X}}/{\Phi _{0}} \approx
10^{-3}$. Theoretically, it is possible to increase the ratio $\pi
{\Phi _{X}}/{\Phi _{0}}$ by using a stronger magnetic field $B_L$
and $B_R$ and/or using soft NAMRs with greater zero-point
fluctuations. However, in practice, the magnetic field is limited to
(upmost) in tens Tesla and the zero-point fluctuations of the NAMRs
is less than $10^{-12}$ m for the most of the experiments.
Therefore, the periodic nature of the Josephson Hamiltonian has no
chance to play a role here. Indeed the situation considered here is
very similar to the optical parametric-down-conversion system,
except for the coefficients of polynomial expansions of the
interaction Hamiltonian.

Now we consider how well a dc-SQUID is approximated by a harmonic
oscillator. Since the barrier of the potential $U\left( \varphi
,\Phi _{X}\right) $ has a finite height, the dynamics of $\varphi $
is not an ideal harmonic oscillator. However, if the energy of the
quasi-particle is small enough, the dynamics of $\varphi $ can be
approximately described by a harmonic oscillator. The maximum number
$N_{\max }$ of energy levels that can be confined in the potential
$U\left( \varphi , 0\right) $ is $N_{\max }\equiv \Delta U/\Omega$,
where the height of the potential is
\begin{equation}
\Delta U=2E_{J}\left[ 2\cos q_{0}+\sin q_{0}\left( 2q_{0}-\pi
\right) \right] \,\,\text{,}
\end{equation}
and $\Omega$ is the frequency of the harmonic oscillator.

\subsection{Hamiltonian of the Coupled System}

Near the minimum potential $U(q_{0},0)$, the free Hamiltonian of the
dc-SQUID can be written as a harmonic oscillator Hamiltonian
\begin{equation}\label{eq:10}
\frac{H_{s}}{\hbar}= E_{c}\,\dot{\varphi}^{2}+ E_{J}^{\prime
}\,\varphi ^{2}\,\,\text{,}
\end{equation}%
with
\begin{eqnarray}
E_{J}^{\prime }=E_{J}\cos q_{0} \,\,\text{,}
\end{eqnarray}
where the constant term has been omitted when we derived the
Hamiltonian in Eq.~(\ref{eq:10}). It is convenient to introduce the
annihilation and creation operators $a$ and $a^{\dagger }$:
\begin{subequations}
\begin{eqnarray}
a &=&\left( \frac{E_{J}^{\prime }}{E_{c}}\right) ^{1/4}\varphi +i\left(
\frac{E_{c}}{E_{J}^{\prime }}\right) ^{1/4}\dot{\varphi}\,\,\text{,} \\
a^{\dagger } &=&\left( \frac{E_{J}^{\prime }}{E_{c}}\right)
^{1/4}\varphi -i\left( \frac{E_{c}}{E_{J}^{\prime }}\right)
^{1/4}\dot{\varphi}\,\,\text{.}
\end{eqnarray}
\end{subequations}
Then the free Hamiltonian in Eq.~(\ref{eq:10}) of the dc-SQUID can
be rewritten in the form
\begin{equation}
H_{s}=\hbar \Omega \, a^{\dagger }\,a\,\,\text{,}
\label{Eq:Hamiltonian SQUID 01}
\end{equation}
with the angular frequency
\begin{eqnarray}
\Omega =\sqrt{E_{c}E_{J}^{\prime }} \,\,\text{.}
\end{eqnarray}
When the energy of the dc-SQUID is not very large ($ \langle
a^{\dagger }a \rangle < N_{\text{max}}$), the dynamics of $\varphi $
is well described by a harmonic oscillator under a suitable bias
magnetic flux threading the loop of the dc-SQUID. It is convenient
to also introduce annihilation and creation operators for the
fundamental flexural modes of the two NAMRs ($i=L,R$)
\begin{subequations}
\label{eq:second quantization of NAMRs}
\begin{eqnarray}
b_{i} &=&\sqrt{\frac{m_{i}\omega _{i}}{2\hbar
}}\,X_{i}+i\sqrt{\frac{1}{2\hbar
m_{i}\omega _{i}}}\,P_{i}\,\,\text{,} \\
b_{i}^{\dagger } &=&\sqrt{\frac{m_{i}\omega _{i}}{2\hbar }}\,X_{i}-i\sqrt{%
\frac{1}{2\hbar m_{i}\omega _{i}}}\,P_{i}\,\,\text{.}
\end{eqnarray}
\end{subequations}
Here, $X_{i}$ and $P_{i}$ are the coordinate and momentum operators
of the $i$th NAMR; $m_{i}$ and $\omega _{i}$ are the effective mass
and angular frequency of the $i$th NAMR. The effective angular
frequency $\omega _{i}$ is not the one of the fundamental flexural
mode, which is modified by the second term in the potential Eq.~(\ref%
{eq:potential}). Then the free Hamiltonian of the two NAMRs can be
written in the form
\begin{equation}
H_{\text{NAMR}}=\hbar \omega _{L}\,b_{L}^{\dagger }b_{L}+\hbar
\omega _{R}\,b_{R}^{\dagger }b_{R}\,\,\text{,}
\label{Eq:Hamiltonian NAMRs 01}
\end{equation}%
where the constant terms have been omitted. Thus, in terms of
creation and annihilation operators, from Eq.~(\ref{eq:potential}),
the \textit{interaction} Hamiltonian between these two NAMRs and the
dc-SQUID are given by
\begin{eqnarray}
V &=&-\left[ g_{L}\left( b_{L}+b_{L}^{\dagger }\right) +g_{R}\left(
b_{R}+b_{R}^{\dagger }\right) \right] ^{2}  \notag \\
&&\left[ c_{1}\left( a+a^{\dag }\right) +c_{2}\left( a+a^{\dag
}\right) ^{2}\right] \,\,\text{,}  \label{eq:coupling 01}
\end{eqnarray}%
where
\begin{subequations}
\label{eq:coefficent}
\begin{eqnarray}
g_{L} &=&\frac{\pi B_{L}l}{\Phi _{0}}\sqrt{\frac{\hbar }{2m_{L}\omega _{L}}}%
\,\,\text{,} \\
g_{R} &=&\frac{\pi B_{R}l}{\Phi _{0}}\sqrt{\frac{\hbar }{2m_{R}\omega _{R}}}%
\,\,\text{,} \\
c_{1} &=&\frac{\Omega}{2} (\tan q_{0})\left( \frac{ E_{J} \cos q_{0}
} {E_{c}}\right) ^{1/4}
\,\,\text{,} \\
c_{2} &=&\frac{\Omega}{8} \,\,\text{.}
\end{eqnarray}
\end{subequations}

The interaction Hamiltonian~(\ref{eq:coupling 01}) is central to
this work. Notice that it contains both linear and nonlinear terms.
Generally, it is very difficult to evaluate the behavior of this
coupled system. However, since the frequency $\Omega $ of the
dc-SQUID can be set by the bias current $I_b$, we can reduce the
interaction Hamiltonian $V$ in Eq.~(\ref{eq:coupling 01}) to a
simplified form by invoking the rotating wave approximation. We now
rewrite the interaction Hamiltonian $V$ in Eq.~(\ref{eq:coupling
01}) in the interaction picture, with the free Hamiltonian
\begin{equation}
H_{0}=\hbar \Omega a^{\dagger }a+\hbar \omega _{L}b_{L}^{\dagger
}b_{L}+\hbar \omega _{R}b_{R}^{\dagger }b_{R}
\end{equation}
Then the terms of the interaction Hamiltonian $V$ can be classified
by the ways that the frequencies $\omega_L$, $\omega_R$, and
$\Omega$ can be combined. In Table~\ref{tab:terms}, we list half of
the coupling terms and the combinations of their frequencies. The
other half are their corresponding Hermitian conjugate terms, which
have the same frequencies but with a negative sign.

In Table~\ref{tab:terms}, it can be seen that, for large detuning,
one needs to mainly consider the zero-frequency terms in the first
row of the table. Then this interaction Hamiltonian $V$ enables a
quantum nondemolition measurement of discrete Fock states of a NAMR,
as discussed in Ref.~\onlinecite{Buks2006oct}. When the frequency of
the dc-SQUID is set at some special value, one can mainly consider
the resonant terms. For example, if the frequency of the dc-SQUID
and those of the two NAMRs are properly set so that $\Omega \neq
\omega _{L}=\omega _{R}$ and also $\Omega \neq 2 \omega_{L}= 2
\omega_{R}$, then only the zero-frequency terms and resonant terms
in the interaction Hamiltonian $V$ are kept under the rotating wave
approximation. The reduced interaction Hamiltonian $V_r$ consists of
the terms
\begin{eqnarray}
V_r=c_{2}g_{L}g_{R}\,b_{R}^{\dag }b_{L}\,a^{\dag }a+\text{H.c.}
\,\,\text{,}
\end{eqnarray}
which in fact offers us a mechanism for coupling two NAMRs. Thus,
our proposed device offers a flexible (literately) model for the
control and measurement of NAMRs.

\begin{table}[tp]
\centering%
\begin{tabular}{|c|c|c|}
\hline\hline ~ & Frequencies & Interaction terms \\
\hline 0 & $0$ & $\frac{1}{2}\,c_{2}( g_{L}^{2}b_{L}^{\dagger
}b_{L}+g_{R}^{2}b_{R}^{\dagger }b_{R}) a^{\dag }a$, \\
1 & $2\omega _{L}$ & $c_{2}g_{L}^{2}b_{L}^{2}a^{\dag }a$, \\
2 & $2\omega _{R}$ & $c_{2}g_{R}^{2}b_{R}^{2}a^{\dag }a$, \\
3 & $\omega _{L}+\omega _{R}$ & $c_{2}g_{L}g_{R}b_{L}b_{R}a^{\dag }a$, \\
4 & $\omega _{L}-\omega _{R}$ & $c_{2}g_{L}g_{R}b_{R}^{\dag }b_{L}a^{\dag }a$%
, \\
5 & $2\Omega $ & $c_{2}( g_{L}^{2}b_{L}^{\dagger
}b_{L}+g_{R}^{2}b_{R}^{\dagger }b_{R}) a^{2}$, \\
6 & $\Omega $ & $c_{1}( g_{L}^{2}b_{L}^{\dagger
}b_{L}+g_{R}^{2}b_{R}^{\dagger }b_{R}) a$, \\
7 & $2\omega _{L}+2\Omega $ & $c_{2}g_{L}^{2}b_{L}^{2}a^{2}$, \\
8 & $2\omega _{L}-2\Omega $ & $c_{2}g_{L}^{2}b_{L}^{2}a^{\dag 2}$, \\
9 & $2\omega _{R}+2\Omega $ & $c_{2}g_{R}^{2}b_{R}^{2}a^{2}$, \\
10 & $2\omega _{R}-2\Omega $ & $c_{2}g_{R}^{2}b_{R}^{2}a^{\dag 2}$, \\
11 & $\omega _{L}+\omega _{R}+2\Omega $ & $c_{2}g_{L}g_{R}b_{L}b_{R}a^{2}$,
\\
12 & $\omega _{L}+\omega _{R}-2\Omega $ & $c_{2}g_{L}g_{R}b_{L}b_{R}a^{\dag
2}$, \\
13 & $\omega _{L}-\omega _{R}+2\Omega $ & $c_{2}g_{L}g_{R}b_{L}b_{R}^{\dag
}a^{2}$, \\
14 & $\omega _{L}-\omega _{R}-2\Omega $ & $c_{2}g_{L}g_{R}b_{L}b_{R}^{\dag
}a^{\dag 2}$, \\
15 & $2\omega _{L}+\Omega $ & $c_{1}g_{L}^{2}b_{L}^{2}a$, \\
16 & $2\omega _{L}-\Omega $ & $c_{1}g_{L}^{2}b_{L}^{2}a^{\dag }$, \\
17 & $2\omega _{R}+\Omega $ & $c_{1}g_{R}^{2}b_{R}^{2}a$, \\
18 & $2\omega _{R}-\Omega $ & $c_{1}g_{R}^{2}b_{R}^{2}a^{\dag }$, \\
19 & $\omega _{L}+\omega _{R}+\Omega $ & $c_{1}g_{L}g_{R}b_{L}b_{R}a$, \\
20 & $\omega _{L}+\omega _{R}-\Omega $ & $c_{1}g_{L}g_{R}b_{L}b_{R}a^{\dag }$%
, \\
21 & $\omega _{L}-\omega _{R}+\Omega $ & $c_{1}g_{L}g_{R}b_{L}b_{R}^{\dag }a$%
, \\
22 & $\omega _{L}-\omega _{R}-\Omega $ & $c_{1}g_{L}g_{R}b_{L}b_{R}^{\dag
}a^{\dag }$, \\ \hline\hline
\end{tabular}%
\caption{Terms in the interaction Hamiltonian Eq.~(\ref{eq:coupling
01}) and their frequencies, in the interaction picture.
\label{tab:terms}}
\end{table}

\section{Two-Mode Squeezed states of two nanomechanical resonators}

\label{sec:squeezing}

In this section we focus on the two-mode squeezed states of the two
NAMRs. It is possible to produce entangled states of the two NAMRs
by considering the analog of the parametric down-conversion in
quantum optics. The zero-frequency terms in Table~\ref{tab:terms}
commute with the free Hamiltonian (\ref{Eq:Hamiltonian SQUID 01}) of
the dc-SQUID and the free Hamiltonian (\ref {Eq:Hamiltonian NAMRs
01}) of the two NAMRs. Let us assume that the proposed circuit works
at low temperature. If the two NAMRs are initially in the vacuum
state or in very low-energy states, then we have $\delta _{LR} \ll
\Omega$, with
\begin{eqnarray}
\delta _{LR}=c_{2}\left( g_{R}^{2}\, \langle b_{R}^{\dagger }b_{R}
\rangle +g_{L}^{2}\,\langle b_{L}^{\dagger }b_{L} \rangle \right)
\,\,\text{.}
\end{eqnarray}
Then we can rewrite the free Hamiltonians of the dc-SQUID
Eq.~(\ref{Eq:Hamiltonian SQUID 01}) and the two NAMRs
Eq.~(\ref{Eq:Hamiltonian NAMRs 01}) as
\begin{equation}
H_{0}^{\prime }=\Omega ^{\prime }a^{\dagger }a+\omega
_{L}b_{L}^{\dagger }b_{L}+\omega _{R}b_{R}^{\dagger
}b_{R}\,\,\text{,}  \label{eq:H0 prime}
\end{equation}%
where $\Omega ^{\prime }=\Omega -\delta _{LR}$. By properly setting
the bias current $I_{b}$ one can let $\Omega ^{\prime }-\omega
_{L}-\omega _{R}=0$. Then, in the interaction picture, after
adopting the rotating wave approximation, we simplify the
interaction Hamiltonian between
two NAMRs and dc-SQUID as%
\begin{equation}
V^{\prime }=\eta \left( a^{\dag }\,b_{L}\,b_{R}+a\,b_{L}^{\dagger
}\,b_{R}^{\dagger }\right) \,\,\text{,} \label{eq:coupling 02}
\end{equation}%
where
\begin{eqnarray}
\eta =-c_{1}g_{L}g_{R} \,\,\text{.}
\end{eqnarray}
Driven by this interaction Hamiltonian $V^{\prime }$,
\textit{two-mode squeezed states of the two NAMRs} can be produced
in the device similarly to a light beam interacting inside a
nonlinear medium in quantum optics, because both of them follow the
\textit{same} Hamiltonian (\ref{eq:coupling 02}).

We now consider that the mode of the dc-SQUID is in a coherent state
$\left\vert \alpha \right\rangle $, where $\left\vert \alpha
\right\vert \gg 1$. Then we can treat the mode of the dc-SQUID as a
classical field and replace the operator $a$ in the Hamiltonian
$V^{\prime }$ in Eq.~(\ref{eq:coupling 02}) by a complex number
$\left\vert \alpha \right\vert \exp \left( -i\phi \right)$. Then, in
the interaction picture defined by the Hamiltonians (\ref{eq:H0
prime}) and (\ref{eq:coupling 02}), the dynamics of the coupled
system is described by the following Hamiltonian
\begin{equation}
V_I=e^{i\phi }\left\vert \alpha \right\vert \eta\,
b_{L}\,b_{R}+e^{-i\phi }\left\vert \alpha \right\vert \eta\,
b_{L}^{\dagger }\,b_{R}^{\dagger }\,\,\text{.} \label{eq:coupling
03}
\end{equation}%
The motions of $b_{L}$ and $b_{R}$ are
\begin{subequations}
\label{eq:ME of NAMRs}
\begin{eqnarray}
b_{L}\left( t\right) &=&\cosh \left( \gamma \right) b_{L}
-ie^{-i\phi }\sinh \left( \gamma \right) b_{R}^{\dagger }
\,\,\text{,} \\
b_{R}\left( t\right) &=&\cosh \left( \gamma \right) b_{R}
-ie^{-i\phi }\sinh \left( \gamma \right) b_{L}^{\dag } \,\,\text{,}
\end{eqnarray}%
\end{subequations}
in the interaction picture of the Hamiltonians~(\ref{eq:H0 prime})
and (\ref{eq:coupling 03}), with
\begin{eqnarray}
\gamma = \left\vert \alpha \right\vert \eta \, t \,\,\text{.}
\end{eqnarray}

The generation of two-mode squeezed states of these two NAMRs can be
shown by their collective coordinate and momentum operators
\begin{subequations}
\begin{eqnarray}
X_{T}\left( t\right)  &=&X_{L}\left( t\right) +X_{R}\left( t\right)
\text{,}
\\
P_{T}\left( t\right)  &=&P_{L}\left( t\right) +P_{R}\left( t\right)
\text{,}
\end{eqnarray}
\end{subequations}
where, $X_{i}\left( t\right) $ and $P_{i}\left( t\right) $, $i=L,R$,
are defined by Eq.~(\ref{eq:second quantization of NAMRs}) by
substituting $b_{i}$ and $b_{i}^{\dagger}$ with $b_{i}(t)$ and
$b_{i}^{\dagger}(t)$ in Eq.~(\ref{eq:ME of NAMRs}). The uncertainty
relation for the collective coordinate and momentum operators
$X_{T}\left( t\right) $ and $P_{T}\left( t\right) $ is
\begin{eqnarray}\label{eq:variance02}
\Delta \left[ X_{T}\left( t\right) \right]\, \Delta \left[
P_{T}\left( t\right) \right]=\hbar\left\vert \cosh
^{2}\gamma+e^{2i\phi }\sinh ^{2}\gamma  \right\vert \text{.}
\end{eqnarray}
In Eq.~(\ref{eq:variance02}) we have assumed that the zero-point
fluctuation of positions of the left NAMR
\begin{eqnarray}
\delta_{L} = \sqrt{{\hbar }/{(2m_{L}\omega _{L})}} ,
\end{eqnarray}
 and that of the right NAMR
\begin{eqnarray}
\delta_{R} = \sqrt{{\hbar }/{(2m_{R}\omega _{R})}}
\end{eqnarray}
are the same. Here
\begin{eqnarray}
\delta _{X}=\sqrt{2}\delta _{L}=\sqrt{2}\delta _{R}
\end{eqnarray}
is defined as the zero-point fluctuation of the collective
coordinates $X_{T}$ of the two NAMRs. And $\zeta _{P}=\sqrt{2}\zeta
_{L}=\sqrt{2}\zeta _{R}$ is defined as the zero-point fluctuation of
the collective momentums $P_{T}$ of the two NAMRs. Here, $\zeta
_{i}^2=\hbar/(2\delta _{i}^2)$, $i=L,R$.

If we choose $\phi =-\pi /2$, then the variance of the collective
coordinates $X_{T}(t)$ becomes
\begin{equation}
\Delta \left[ X_{T}(t)\right] =\delta _{X} \, \exp(\gamma
)\,\,\text{.} \label{eq:variance-ideal}
\end{equation}%
Notice that $\gamma <0$ because $\gamma=-c_1 g_L g_R |\alpha| \, t$.
Therefore, perfect two-mode squeezed states, i.e., pure entangled
states, of the two NAMRs are generated.

The variance of $X_{T}(t)$ (the entanglement) was obtained above by
assuming that both the left and right NAMRs be in their ground
states. It can be checked that if both the left and right NAMRs are
initially in coherent states or thermal states, then the
Hamiltonian~(\ref{eq:coupling 03}) will not produce entangled states
of them. However, if only one of the NAMRs is initially prepared
into a number state, then entangled states of these two NAMRs can be
generated by the Hamiltonian~(\ref{eq:coupling 03}). For example,
when the left and the right NAMRs are initially prepared in the
number states $\vert 0 \rangle $ and $\vert 1 \rangle $,
respectively, then the Bell-type entangled state $a_1 \vert 01
\rangle + a_2 \vert 10 \rangle$ can be generated. Here, $a_1$ and
$a_2$ are complex numbers. When one of the two NAMRs is initially in
a coherent state and the other one is in the vacuum state, then the
so-called ``single-photon-added coherent states''~\cite{Zavatta2004}
can be generated by the Hamiltonian~(\ref{eq:coupling 03}).

\begin{figure}[tp]
\centering
\includegraphics[bb= 70 260 560 540, clip, width=8cm]{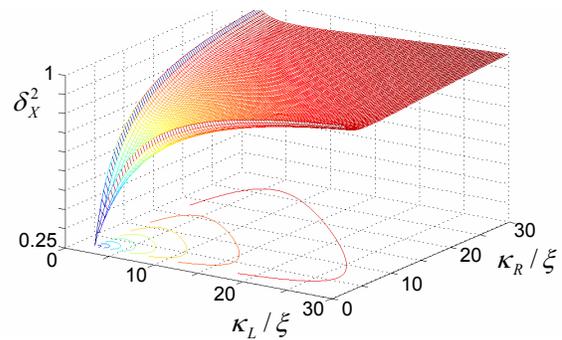}
\caption{(Color online) The squared variance of the collective
coordinates $X_{T}$ of the two NAMRs as the function of damping
rates $\kappa_L$ and $\kappa_R$ of the left and right NAMRs, both
$\kappa_L$ and $\kappa_R$ are normalized by the effective coupling
constant $\xi$. Here, $\delta _{X}$ is the zero fluctuation of the
collective coordinates of the two NAMRs.} \label{fig:variance}
\end{figure}

Let us now consider the more realistic case where both the dc-SQUID
and the two NAMRs are coupled to their environments. The quality
factors of the two NAMRs with GHz frequency are smaller than that of
the dc-SQUID~\cite{Gaidarzhy2005,Day2003}. The quality factor of a
GHz NAMR is of the order of $10^3$, while that of a superconducting
circuit can be as large as $10^6$. Therefore, below we consider the
noise from the environment acting on the two NAMRs. To include
damping effects, due to the noise from the environments, on the
dynamics of the two NAMRs, we use the Heisenberg-Langevin equation
method~\cite{Scullybook1997}. Then, for the motions of the operators
of the NAMRs, we have the following set of equations:
\begin{subequations}
\begin{eqnarray}
\frac{d}{dt}b_{L} &=&-\xi b_{R}^{\dagger }-\frac{\kappa _{L}}{2}%
b_{L}+F_{L}\left( t\right) \,\,\text{,} \\
\frac{d}{dt}b_{R} &=&-\xi b_{L}^{\dag }-\frac{\kappa _{R}}{2}%
b_{R}+F_{R}\left( t\right) \,\,\text{,}
\end{eqnarray} \label{eq:Heisenberg-Langevin 01}
\end{subequations}
As in the ideal case in Eq.~(\ref{eq:variance-ideal}) we also let
$\phi =-\pi /2$. Here,
\begin{eqnarray}
\xi=|\alpha|\,\eta
\end{eqnarray}
is the \textit{effective coupling strength} between the dc-SQUID and
two NAMRs. Also, $\kappa _{L}$ and $\kappa _{R}$ represent the
damping rates of the left and right NAMRs, respectively; and the
associated noise operators are $F_{L}\left( t\right) $ and
$F_{R}\left( t\right) $. We evaluate the properties of the states of
the two NAMRs by the variance of the collective coordinates $X_{T}$.
We find that the damping of the two NAMRs help producing two-mode
squeezed states of the two NAMRs, regardless of the initial states.
The variance of the collective coordinates $X_{T}$ is calculated as
\begin{eqnarray}
[\Delta \left( X_{T}\right)]^2  &=&\frac{\delta _{L}^{2}}{\kappa
_{+}}\left( 2\kappa _{R}\,\Delta _{ \xi }+\kappa _{-}\right) +
\frac{\delta _{R}^{2}}{\kappa _{+}}\left( 2\kappa _{L}\,\Delta _{\xi
}-\kappa _{-}\right)   \notag \\
&&-8\frac{\delta _{L}\,\delta _{R}}{\kappa _{+}}\,\xi \,\Delta _{\xi }\text{%
,}
\end{eqnarray}%
under the Markov approximation and in the overdamped case: $\xi <
\kappa_{L}/2$ and $\xi < \kappa_{R}/2$. In the Appendix A we outline
the main ideas of the derivation. Here,
\begin{eqnarray}
\kappa _{\pm } =\kappa _{L}\pm \kappa _{R}\,\text{,}\,\, \Delta
_{\xi } =\frac{\kappa _{L}\kappa _{R}}{\kappa _{L}\kappa _{R}-4\xi
^{2}}\,\,\text{.}
\end{eqnarray}%
When the zero-point fluctuations of the left and right NAMRs are
equal, we have
\begin{equation}
[\Delta \left( X_{T}\right)]^2 =\frac{\delta _{X}^{2}}{4}\Delta _{\xi }\left( 1-%
\frac{4\xi }{\kappa _{L}+\kappa _{R}}\right) \,\,\text{.}
\end{equation}%
Fig.~\ref{fig:variance} shows, the variance of the collective
coordinates $X_{T}$ versus the damping rates $\kappa_{L}$ and
$\kappa_{R}$. It is clear from Fig.~\ref{fig:variance} that
appreciable squeezing can be generated even when the dampings of the
two NAMRs are severe (ten times the coupling constant $\xi$). This
indicates that the squeezing is robust against damping. The maximum
squeezing is obtained when both damping rates (for the left and
right NAMRs) approach the coupling strength between them and the
dc-SQUID. Since the coupling strength $\xi$ is proportion to $\left
\vert \alpha \right\vert $, one can increase the squeezing rate by
gradually increasing the power of the microwave applied to the
dc-SQUID. As the damping rates of the two NAMRs increase, the
squeezing effect decreases steadily.

To consider the experimental feasibility of our proposal, we choose
the following parameters for the two NAMRs and the dc-SQUID
\begin{subequations}
\begin{eqnarray}
m_{L} &=&m_{R}=10^{-18}\text{ Kg,} \\
\omega _{L} &=&1.5 \text{ GHz,} \\
\omega _{R} &=&1.2 \text{ GHz,} \\
l &=&10 \text{ }\mu \text{m,} \\
B_{L} &=&B_{R}=1\text{ T,} \\
\kappa _{L} &=&\kappa _{R}=2 \text{ MHz,} \\
E_{C} &=&0.061 \text{ GHz,} \\
E_{J} &=&120 \text{ GHz.}
\end{eqnarray}
\end{subequations}
It was already demonstrated in
experiments~\cite{Gaidarzhy2005,Gaidarzhy2005June} that a 10 $\mu$m
long doubly-clamped beam can oscillate with a frequency of several
Giga Hertzs. The effective mass of this antenna-shaped beam is much
smaller than its weight. And its effective mass can be further
modified when the beams are under strains and stresses. The numbers
used here for the dc-SQUID are also consistent with the experimental
numbers shown in Ref.~\onlinecite{Valenzuela2006Dec}. Then the
additional magnetic flux from the two NAMRs, $\Phi_X/\Phi_0 \simeq 5
\times 10^{-3} \ll 1$, would satisfy our assumption in
section~\ref{sec:the device}. The maximum number $N_{\text{max}}$ of
energy levels confined in the current biased potential energy is
calculated as $N_{\text{max}} \simeq 150$. Therefore, the harmonic
oscillator approximation and the classical field approximation for
the dc-SQUID are both possible. The time needed to obtain the
two-mode squeezed state is determined by the effective coupling
constant $\xi$. Assuming the same damping rates for the two NAMRs,
the maximum squeezing $\Delta \left( X_{T}\right)=(1/2)\,\delta_X$
can be obtained. Therefore, it should be possible to realize our
proposal of generating two-mode squeezed states of the two NAMRs
with current experimental conditions.

To experimentally detect the generated two-mode squeezed state of
the two NAMRs, a (in principle) relatively direct method would be
checking the variance of the collective coordinate $X_{T}$ of the
two NAMRs. Since the left and right NAMRs are symmetric in the
interaction Hamiltonian (\ref{eq:coupling 01}), they can be treated
as one virtual NAMR. To detect two-mode squeezed states of the two
NAMRs, the detection methods should be able to approach standard
quantum limit of the NAMRs. With traditional displacement detection
methods~\cite{Clelandbook2002,Schwab2005PT,Munday2005,Munday2006,Capasso2007},
such as optical interferences, magnetic-motive method and coupled
single electron transistor, the best record of detection precision
was about 4.3 standard quantum limits~\cite{LaHaye2004}. There are
also other proposals for displacement detection by coupling the
mechanical oscillator to some two level
system~\cite{Armour2002,Zhou2006Dec,Xue2007NJP}. These methods in
principle can detect quantum states of the NAMRs. After the
entangled state is generated, one can switch the dc-SQUID to the
phase qubit regime~\cite{Cooper2004}, and utilize the nonlinear
coupling between the virtual NAMR and the dc-SQUID. Then the SQUID
can be used to measure the variance of $X_{T}$, as discussed in
Ref.~\onlinecite{Zhou2006Dec}.

\section{Discussions and conclusions}

\label{sec:discussions}

In our proposal, we only consider the fundamental vibration modes of
the NAMRs. Generally, there are also vibration modes with higher
frequencies, torsional and strain-stress oscillations in the
NAMRs~\cite{Clelandbook2002}. The vibration modes with higher
frequencies will be excited only when they happen to resonate with
the dc-SQUID, which can be easily avoided by optimizing the
parameters of the NAMRs. As for torsional and strain-stress
oscillations, they are hardly coupled to the dc-SQUID. These modes
of oscillations of the NAMRs will not change the magnetic flux
through the dc-SQUID, thereby these modes cannot be coupled to the
dc-SQUID, even through their frequencies match the resonant
condition. It is similar to the case of the experiments of
magnetomotive detection of flexural oscillation of
NAMRs~\cite{Clelandbook2002}, where torsional and strain-stress
oscillations have been neglected.

As mentioned in Section.~\ref{sec:squeezing}, to generate two-mode
squeezed states of NAMRs, the NAMRs should start in their ground
states or number states. The NAMRs should be cooled such that the
thermal excitation energy is less than those corresponding to the
NAMRs' frequencies. For a one-GHz NAMR, this means that the
temperature should be below 50 mK, which is still within the
capability of dilution refrigerators. In principle it is possible to
prepare the NAMRs used in our proposal in their ground states.
Moreover, recently there have been many efforts in reducing the
temperature of mechanical resonators by active
cooling~\cite{Naik2006,Schliesser2006,Kleckner2006,Gigan2006,Poggio2007}.
Also a temperature as low as 5 mK was already demonstrated for a
mechanical resonator~\cite{Poggio2007}. Besides, there are also
theoretical proposals for the production of number states of
NMARs~\cite{Irish2003}. Therefore, though currently the ground
states and (or) number states of NAMRs might be difficult to prepare
experimentally, we expect these to be realized more earlier in the
near future. For example, if the temperature of a GHz NAMR reaches
$10$ mk, the thermal occupation number will be $\sim 10^{-3}$, which
is nearly a ground state.

In conclusion, we have proposed a device to couple a dc-SQUID to two
NAMRs, which can be used to create an effective coupling between
these two NAMRs, and also to measure and control the two NAMRs. We
have shown that two-mode squeezed states can be generated in a
robust fashion by this device, in analogy to the two-mode parametric
down-conversion process in quantum optics. This two-mode
down-conversion process offers us a protocol of producing
entanglement in two mechanical resonators in a solid state device,
while previous proposals, see, e.g.,
Refs.~\onlinecite{Mancini2002,Zhang2003,Pirandola2006}, were based
on entanglement-swapping by the assistance of photons. Our proposal
might be promising for the experimental test of the existence of
entangled states of macroscopic objects.

\section{Acknowledgement}

FN was supported in part by the US National Security Agency (NSA),
Army Research Office (ARO), Laboratory of Physical Sciences (LPS),
and the National Science Foundation grant No. EIA-0130383. CPS was
supported in part by the NSFC with grant Nos. 90203018, 10474104 and
60433050; and the National Fundamental Research Program of China
with Nos. 2001CB309310 and 2005CB724508.

\appendix

\section{Heisenberg-Langevin Equation for Two Nanomechanical resonators}

We consider the Markov approximation $\left\langle
F_{L,R}\right\rangle =\left\langle F_{L,R}^{\dag }\right\rangle =0$.
Using Eqs.~(\ref{eq:Heisenberg-Langevin
01}a)-(\ref{eq:Heisenberg-Langevin 01}b), a solution of the
expectation values $\left\langle b_{L}\left( t\right) \right\rangle
$ and $\left\langle b_{R}\left( t\right) \right\rangle $ can be
given as
\begin{subequations}
\begin{eqnarray}
\left\langle b_{L}\right\rangle &=&e^{-\frac{1}{2}\kappa
_{L}t}\left[ b_{L}\left( 0\right) \cosh \left( \xi t\right)
-b_{R}^{\dagger }\left(
0\right) \sinh \left( \xi t\right) \right] \,\,\text{,}  \notag \\
&& \\
\left\langle b_{R}\right\rangle &=&e^{-\frac{1}{2}\kappa
_{R}t}\left[ b_{R}\left( 0\right) \cosh \left( \xi t\right)
-b_{L}^{\dag }\left( 0\right)
\sinh \left( \xi t\right) \right] \,\,\text{.}  \notag \\
&&
\end{eqnarray}
\end{subequations}%
It is seen that below the thresholds $\xi <\kappa _{L}/2$ and $\xi <
\kappa _{R}/2$ we have
\begin{equation}
\left\langle b_{L}\right\rangle =\left\langle b_{R}\right\rangle
=0\,\,\text{.}
\end{equation}%
The variance of the collective coordinates $X_{T}$ can be evaluated
by the expectation values of the bilinear operators of the two
NAMRs. These are the expectation values of the quadratic operators
of the left NAMR
\begin{subequations}
\begin{eqnarray}
L_{1} &=&\left\langle b_{L}^{2}\right\rangle \text{, }L_{3}=\left\langle
b_{L}^{\dag 2}\right\rangle \,\,\text{,} \\
L_{2} &=&\left\langle b_{L}^{\dag }b_{L}+b_{L}b_{L}^{\dag
}\right\rangle \,\,\text{,}
\end{eqnarray}
\end{subequations}%
the expectation values of the quadratic operators of the right NAMR
\begin{subequations}
\begin{eqnarray}
R_{1} &=&\left\langle b_{R}^{2}\right\rangle \text{, }R_{3}=\left\langle
b_{R}^{\dag 2}\right\rangle \,\,\text{,} \\
R_{2} &=&\left\langle b_{R}^{\dag }b_{R}+b_{R}b_{R}^{\dag
}\right\rangle \,\,\text{,}
\end{eqnarray}
\end{subequations}%
and the expectation values of the quadratic operators of both NAMRs
\begin{subequations}
\begin{eqnarray}
C_{1} &=&\left\langle b_{L}b_{R}+b_{R}b_{L}\right\rangle = C_{4}^{\dagger}\,\,\text{,} \\
C_{2} &=&\left\langle b_{L}b_{R}^{\dagger }+b_{R}^{\dagger
}b_{L}\right\rangle = C_{3}^{\dagger} \,\,\text{.}
\end{eqnarray}
\end{subequations}

From Eq.~(\ref{eq:Heisenberg-Langevin 01}) it is found that these
expectation values satisfy a closed set of equations of
motion~\cite{Scullybook1997}. To determine the values involving the
expectation values of the products of the noise operators and the
operators of the NAMRs, we rewrite Eq.~(\ref{eq:Heisenberg-Langevin
01}a)-(\ref{eq:Heisenberg-Langevin 01}b) and their corresponding
Hermitian ones in the matrix form
\begin{equation}
\mathcal{\dot{B}=-MB+F}\,\,\text{,}  \label{eq:matrix form of H-L}
\end{equation}%
where $\mathcal{B=}\left[ b_{L}\left( t\right) ,b_{L}^{\dag }\left( t\right)
,b_{R}\left( t\right) ,b_{R}^{\dag }\left( t\right) \right] ^{T}$ and $\mathcal{%
F}=\left[ F_{L}\left( t\right) ,F_{L}^{\dag }\left( t\right) ,F_{R}\left(
t\right) ,F_{R}^{\dag }\left( t\right) \right] ^{T}$ are vectors, and $\left[ ...%
\right] ^{T}$ represents the transpose operation. Here
\begin{equation}
\mathcal{M=}\left[
\begin{array}{cccc}
\frac{\kappa _{L}}{2} & 0 & 0 & \xi \\
0 & \frac{\kappa _{L}}{2} & \xi & 0 \\
0 & \xi & \frac{\kappa _{R}}{2} & 0 \\
\xi & 0 & 0 & \frac{\kappa _{R}}{2}%
\end{array}%
\right]  \,\,\text{.}
\end{equation}%
A formal solution of Eq.~(\ref{eq:matrix form of H-L}) is given by
\begin{equation}
\mathcal{B}\left( t\right) =e^{-\mathcal{M}t}\,\mathcal{B}\left(
0\right) +\int_{0}^{t}e^{-\mathcal{M}\left( t-t^{\prime }\right)
}\mathcal{F}\left( t^{\prime }\right) dt^{\prime }\,\,\text{.}
\label{eq:matrix form of H-L 01}
\end{equation}%
Multiplying the above equation by $\mathcal{F}^{\dag }\left( t\right) $ from
the right side, we obtain%
\begin{eqnarray}
&&\mathcal{B}\left( t\right) \mathcal{F}^{\dag }\left( t\right) =e^{-%
\mathcal{M}t}\mathcal{B}\left( 0\right) \mathcal{F}^{\dag }\left( t\right)
\notag \\
&&+\int_{0}^{t}e^{-\mathcal{M}\left( t-t^{\prime }\right)
}\mathcal{F}\left( t^{\prime }\right) dt^{\prime }\mathcal{F}^{\dag
}\left( t\right) \,\,\text{.} \label{eq:matrix form of H-L 02}
\end{eqnarray}%
Since the operators of the NAMRs at the initial time $t=0$ are
statistically independent of the noise operators, we have
$\left\langle \mathcal{B}\left( 0\right) \mathcal{F}^{\dag }\left(
t\right) \right\rangle =0$. Using the fact that the corresponding
elements of the matrix of the left part of Eq.~(\ref{eq:matrix form
of H-L 02}) and those of the matrix of the right part of
Eq.~(\ref{eq:matrix form of H-L 02}) are equal, and combining the
Markov approximation, we obtain
\begin{subequations}
\begin{eqnarray}
\left\langle b_{L}\left( t\right) F_{L}^{\dag }\left( t\right) \right\rangle
&=&\frac{\kappa _{L}}{2}\text{, } \\
\left\langle b_{R}\left( t\right) F_{R}^{\dag }\left( t\right)
\right\rangle &=&\frac{\kappa _{R}}{2}\,\,\text{.}
\end{eqnarray}
\end{subequations}
All other products of the operators of the two NAMRs and the noise
operators are zero. Therefore, in the interaction picture, finally
when the expectation values of these bilinear operators do not
change with time, the solution of the above set of equations reads
\begin{eqnarray}
L_{1}=L_{3}=R_{1}=R_{3}=C_{2}=C_{3}=0\,\,\text{,}
\end{eqnarray}
and
\begin{subequations}
\begin{eqnarray}
L_{2} &=&\frac{\kappa _{-}}{\kappa _{+}}+\frac{2\kappa _{R}}{\kappa _{+}}%
\Delta _{\xi }\,\,\text{,} \\
R_{2} &=&-\frac{\kappa _{-}}{\kappa _{+}}+\frac{2\kappa _{L}}{\kappa _{+}}%
\Delta _{\xi }\,\,\text{,} \\
C_{1} &=&C_{4}=-\frac{4\xi }{\kappa _{+}}\Delta _{\xi }\,\,\text{.}
\end{eqnarray}
\end{subequations}
with $\xi <\kappa _{L}/2$ and $\xi < \kappa _{R}/2$. Also, in the
interaction picture, we have $\left\langle b_{L}\left( t\right)
\right\rangle = \left\langle b_{R}\left( t\right) \right\rangle =0$
after a sufficiently long time. Therefore, the variance of the
collective coordinate $X_T$ becomes
\begin{equation}
[\Delta \left( X_{T}\right)]^2 =\delta _{L}^{2}\,L_{2}+\delta
_{R}^{2}\,R_{2}+\delta _{L}\,\delta _{R}\left( C_{1}+C_{4}\right)
\end{equation}
This provides the main result of section \ref{sec:squeezing}.


\begin{thebibliography}{99}
\expandafter\ifx\csname
natexlab\endcsname\relax\def\natexlab#1{#1}\fi
\expandafter\ifx\csname bibnamefont\endcsname\relax
  \def\bibnamefont#1{#1}\fi
\expandafter\ifx\csname bibfnamefont\endcsname\relax
  \def\bibfnamefont#1{#1}\fi
\expandafter\ifx\csname citenamefont\endcsname\relax
  \def\citenamefont#1{#1}\fi
\expandafter\ifx\csname url\endcsname\relax
  \def\url#1{\texttt{#1}}\fi
\expandafter\ifx\csname urlprefix\endcsname\relax\def\urlprefix{URL
}\fi \providecommand{\bibinfo}[2]{#2}
\providecommand{\eprint}[2][]{\url{#2}}

\bibitem[{\citenamefont{Cleland}(2002)}]{Clelandbook2002}
\bibinfo{author}{\bibfnamefont{A.~N.} \bibnamefont{Cleland}},
  \emph{\bibinfo{title}{Foundations of nanomechanics: From Solid-State Theory
  to Device Applications}} (\bibinfo{publisher}{Springer-Verlag},
  \bibinfo{address}{Berlin}, \bibinfo{year}{2002}).

\bibitem[{\citenamefont{Blick et~al.}(2002)\citenamefont{Blick, Erbe, Pescini,
  Kraus, Scheible, Beil, Hoehberger, Hoerner, Kirschbaum, Lorenz
  et~al.}}]{Blick2002}
\bibinfo{author}{\bibfnamefont{R.~H.} \bibnamefont{Blick}},
  \bibinfo{author}{\bibfnamefont{A.}~\bibnamefont{Erbe}},
  \bibinfo{author}{\bibfnamefont{L.}~\bibnamefont{Pescini}},
  \bibinfo{author}{\bibfnamefont{A.}~\bibnamefont{Kraus}},
  \bibinfo{author}{\bibfnamefont{D.~V.} \bibnamefont{Scheible}},
  \bibinfo{author}{\bibfnamefont{F.~W.} \bibnamefont{Beil}},
  \bibinfo{author}{\bibfnamefont{E.}~\bibnamefont{Hoehberger}},
  \bibinfo{author}{\bibfnamefont{A.}~\bibnamefont{Hoerner}},
  \bibinfo{author}{\bibfnamefont{J.}~\bibnamefont{Kirschbaum}},
  \bibinfo{author}{\bibfnamefont{H.}~\bibnamefont{Lorenz}},
  \bibnamefont{et~al.}, \bibinfo{journal}{J. of Phys.: Cond. Mat.}
  \textbf{\bibinfo{volume}{14}}, \bibinfo{pages}{R905} (\bibinfo{year}{2002}).

\bibitem[{\citenamefont{Blencowe}(2004)}]{Blencowe2004}
\bibinfo{author}{\bibfnamefont{M.}~\bibnamefont{Blencowe}},
  \bibinfo{journal}{Phys. Rep.} \textbf{\bibinfo{volume}{395}},
  \bibinfo{pages}{159} (\bibinfo{year}{2004}).

\bibitem[{\citenamefont{Schwab and Roukes}(2005)}]{Schwab2005PT}
\bibinfo{author}{\bibfnamefont{K.}~\bibnamefont{Schwab}} \bibnamefont{and}
  \bibinfo{author}{\bibfnamefont{M.}~\bibnamefont{Roukes}},
  \bibinfo{journal}{Physics Today} \textbf{\bibinfo{volume}{58}}~(6),
  \bibinfo{pages}{36} (\bibinfo{year}{2005}).

\bibitem[{\citenamefont{Huang et~al.}(2003)\citenamefont{Huang, Zorman,
  Mehregany, and Roukes}}]{Huang2003}
\bibinfo{author}{\bibfnamefont{X.~M.~H.} \bibnamefont{Huang}},
  \bibinfo{author}{\bibfnamefont{C.~A.} \bibnamefont{Zorman}},
  \bibinfo{author}{\bibfnamefont{M.}~\bibnamefont{Mehregany}},
  \bibnamefont{and} \bibinfo{author}{\bibfnamefont{M.~L.}
  \bibnamefont{Roukes}}, \bibinfo{journal}{Nature}
  \textbf{\bibinfo{volume}{421}}, \bibinfo{pages}{496} (\bibinfo{year}{2003}).

\bibitem[{\citenamefont{Gaidarzhy
  et~al.}(2005{\natexlab{a}})\citenamefont{Gaidarzhy, Zolfagharkhani, Badzey,
  and Mohanty}}]{Gaidarzhy2005}
\bibinfo{author}{\bibfnamefont{A.}~\bibnamefont{Gaidarzhy}},
  \bibinfo{author}{\bibfnamefont{G.}~\bibnamefont{Zolfagharkhani}},
  \bibinfo{author}{\bibfnamefont{R.~L.} \bibnamefont{Badzey}},
  \bibnamefont{and} \bibinfo{author}{\bibfnamefont{P.}~\bibnamefont{Mohanty}},
  \bibinfo{journal}{Phys. Rev. Lett.} \textbf{\bibinfo{volume}{94}},
  \bibinfo{pages}{030402} (\bibinfo{year}{2005}{\natexlab{a}}).

\bibitem[{\citenamefont{Schwab et~al.}(2005)\citenamefont{Schwab, Blencowe,
  Roukes, Cleland, Girvin, Milburn, and Ekinci}}]{Schwab2005comment}
\bibinfo{author}{\bibfnamefont{K.~C.} \bibnamefont{Schwab}},
  \bibinfo{author}{\bibfnamefont{M.~P.} \bibnamefont{Blencowe}},
  \bibinfo{author}{\bibfnamefont{M.~L.} \bibnamefont{Roukes}},
  \bibinfo{author}{\bibfnamefont{A.~N.} \bibnamefont{Cleland}},
  \bibinfo{author}{\bibfnamefont{S.~M.} \bibnamefont{Girvin}},
  \bibinfo{author}{\bibfnamefont{G.~J.} \bibnamefont{Milburn}},
  \bibnamefont{and} \bibinfo{author}{\bibfnamefont{K.~L.}
  \bibnamefont{Ekinci}}, \bibinfo{journal}{Phys. Rev. Lett.}
  \textbf{\bibinfo{volume}{95}}, \bibinfo{pages}{248901}
  (\bibinfo{year}{2005}).

\bibitem[{\citenamefont{Gaidarzhy
  et~al.}(2005{\natexlab{b}})\citenamefont{Gaidarzhy, Zolfagharkhani, Badzey,
  and Mohanty}}]{Gaidarzhy2005reply}
\bibinfo{author}{\bibfnamefont{A.}~\bibnamefont{Gaidarzhy}},
  \bibinfo{author}{\bibfnamefont{G.}~\bibnamefont{Zolfagharkhani}},
  \bibinfo{author}{\bibfnamefont{R.~L.} \bibnamefont{Badzey}},
  \bibnamefont{and} \bibinfo{author}{\bibfnamefont{P.}~\bibnamefont{Mohanty}},
  \bibinfo{journal}{Phys. Rev. Lett.} \textbf{\bibinfo{volume}{95}},
  \bibinfo{pages}{248902} (\bibinfo{year}{2005}{\natexlab{b}}).

\bibitem[{\citenamefont{Savel'ev
  et~al.}(2006{\natexlab{a}})\citenamefont{Savel'ev, Hu, and
  Nori}}]{Savel'ev2006NJP}
\bibinfo{author}{\bibfnamefont{S.}~\bibnamefont{Savel'ev}},
  \bibinfo{author}{\bibfnamefont{X.}~\bibnamefont{Hu}}, \bibnamefont{and}
  \bibinfo{author}{\bibfnamefont{F.}~\bibnamefont{Nori}}, \bibinfo{journal}{New
  J. of Phys.} \textbf{\bibinfo{volume}{8}}, \bibinfo{pages}{105}
  (\bibinfo{year}{2006}{\natexlab{a}}); cond-mat/0601019.

\bibitem[{\citenamefont{Savel'ev and Nori}(2004)}]{Savel'ev2004Dec}
\bibinfo{author}{\bibfnamefont{S.}~\bibnamefont{Savel'ev}} \bibnamefont{and}
  \bibinfo{author}{\bibfnamefont{F.}~\bibnamefont{Nori}},
  \bibinfo{journal}{Phys. Rev. B} \textbf{\bibinfo{volume}{70}},
  \bibinfo{pages}{214415} (\bibinfo{year}{2004}).

\bibitem[{\citenamefont{Nishiguchi}(2003)}]{Nishiguchi2003Sep}
\bibinfo{author}{\bibfnamefont{N.}~\bibnamefont{Nishiguchi}},
  \bibinfo{journal}{Phys. Rev. B} \textbf{\bibinfo{volume}{68}},
  \bibinfo{pages}{121305(R)} (\bibinfo{year}{2003}).

\bibitem{Savel'ev2007}
\bibinfo{author}{\bibfnamefont{S.}~\bibnamefont{Savel'ev}},
  \bibinfo{author}{\bibfnamefont{A.~L.}~\bibnamefont{Rakhmanov}},
  \bibinfo{author}{\bibfnamefont{X.}~\bibnamefont{Hu}},
  \bibinfo{author}{\bibfnamefont{A.}~\bibnamefont{Kasumov}}, \bibnamefont{and}
  \bibinfo{author}{\bibfnamefont{F.}~\bibnamefont{Nori}},
  \bibinfo{journal}{Phys. Rev. B}
  \textbf{\bibinfo{volume}{75}}, \bibinfo{pages}{165417}
  (\bibinfo{year}{2007}).

\bibitem[{\citenamefont{Armour et~al.}(2002)\citenamefont{Armour, Blencowe, and
  Schwab}}]{Armour2002}
\bibinfo{author}{\bibfnamefont{A.~D.} \bibnamefont{Armour}},
  \bibinfo{author}{\bibfnamefont{M.~P.} \bibnamefont{Blencowe}},
  \bibnamefont{and} \bibinfo{author}{\bibfnamefont{K.~C.}
  \bibnamefont{Schwab}}, \bibinfo{journal}{Phys. Rev. Lett.}
  \textbf{\bibinfo{volume}{88}}, \bibinfo{pages}{148301}
  (\bibinfo{year}{2002}).

\bibitem[{\citenamefont{Wang et~al.}(2004)\citenamefont{Wang, Gao, and
  Sun}}]{Wang2004}
\bibinfo{author}{\bibfnamefont{Y. D.}~\bibnamefont{Wang}},
  \bibinfo{author}{\bibfnamefont{Y. B.}~\bibnamefont{Gao}}, \bibnamefont{and}
  \bibinfo{author}{\bibfnamefont{C. P.}~\bibnamefont{Sun}}, \bibinfo{journal}{Eur.
  Phys. J. B} \textbf{\bibinfo{volume}{40}}, \bibinfo{pages}{321}
  (\bibinfo{year}{2004}).

\bibitem[{\citenamefont{Zhang et~al.}(2005)\citenamefont{Zhang, Wang, and
  Sun}}]{Zhang2005}
\bibinfo{author}{\bibfnamefont{P.}~\bibnamefont{Zhang}},
  \bibinfo{author}{\bibfnamefont{Y.~D.} \bibnamefont{Wang}}, \bibnamefont{and}
  \bibinfo{author}{\bibfnamefont{C.~P.} \bibnamefont{Sun}},
  \bibinfo{journal}{Phys. Rev. Lett.} \textbf{\bibinfo{volume}{95}},
  \bibinfo{pages}{097204} (\bibinfo{year}{2005}).

\bibitem[{\citenamefont{Sun et~al.}(2006)\citenamefont{Sun, Wei, Liu, and
  Nori}}]{Sun2006}
\bibinfo{author}{\bibfnamefont{C.~P.} \bibnamefont{Sun}},
  \bibinfo{author}{\bibfnamefont{L.~F.} \bibnamefont{Wei}},
  \bibinfo{author}{\bibfnamefont{Y.-X.} \bibnamefont{Liu}}, \bibnamefont{and}
  \bibinfo{author}{\bibfnamefont{F.}~\bibnamefont{Nori}},
  \bibinfo{journal}{Phys. Rev. A} \textbf{\bibinfo{volume}{73}},
  \bibinfo{pages}{022318} (\bibinfo{year}{2006}).

\bibitem[{\citenamefont{Wei et~al.}(2006)\citenamefont{Wei, Liu, Sun, and
  Nori}}]{Wei2006}
\bibinfo{author}{\bibfnamefont{L.~F.} \bibnamefont{Wei}},
  \bibinfo{author}{\bibfnamefont{Y.-X.} \bibnamefont{Liu}},
  \bibinfo{author}{\bibfnamefont{C.~P.} \bibnamefont{Sun}}, \bibnamefont{and}
  \bibinfo{author}{\bibfnamefont{F.}~\bibnamefont{Nori}},
  \bibinfo{journal}{Phys. Rev. Lett.} \textbf{\bibinfo{volume}{97}},
  \bibinfo{pages}{237201} (\bibinfo{year}{2006}).

\bibitem[{\citenamefont{Buks et~al.}(2006)\citenamefont{Buks, Arbel-Segev,
  Zaitsev, Abdo, and Blencowe}}]{Buks2006oct}
\bibinfo{author}{\bibfnamefont{E.}~\bibnamefont{Buks}},
  \bibinfo{author}{\bibfnamefont{E.}~\bibnamefont{Arbel-Segev}},
  \bibinfo{author}{\bibfnamefont{S.}~\bibnamefont{Zaitsev}},
  \bibinfo{author}{\bibfnamefont{B.}~\bibnamefont{Abdo}}, \bibnamefont{and}
  \bibinfo{author}{\bibfnamefont{M.~P.} \bibnamefont{Blencowe}},
  \bibinfo{journal}{quant-ph/0610158}  (\bibinfo{year}{2006}).

\bibitem[{\citenamefont{Xue et~al.}(2007)\citenamefont{Xue, Wang, Sun, Okamoto,
  Yamaguchi, and Semba}}]{Xue2007NJP}
\bibinfo{author}{\bibfnamefont{F.}~\bibnamefont{Xue}},
  \bibinfo{author}{\bibfnamefont{Y.~D.} \bibnamefont{Wang}},
  \bibinfo{author}{\bibfnamefont{C.~P.} \bibnamefont{Sun}},
  \bibinfo{author}{\bibfnamefont{H.}~\bibnamefont{Okamoto}},
  \bibinfo{author}{\bibfnamefont{H.}~\bibnamefont{Yamaguchi}},
  \bibnamefont{and} \bibinfo{author}{\bibfnamefont{K.}~\bibnamefont{Semba}},
  \bibinfo{journal}{New Journal of Physics} \textbf{\bibinfo{volume}{9}},
  \bibinfo{pages}{35} (\bibinfo{year}{2007}).

\bibitem[{\citenamefont{Xue et~al.}(2007)\citenamefont{Xue, Zhong, Li, and
  Sun}}]{Xue2007prb}
\bibinfo{author}{\bibfnamefont{F.}~\bibnamefont{Xue}},
  \bibinfo{author}{\bibfnamefont{L.}~\bibnamefont{Zhong}},
  \bibinfo{author}{\bibfnamefont{Y.}~\bibnamefont{Li}}, \bibnamefont{and}
  \bibinfo{author}{\bibfnamefont{C.~P.} \bibnamefont{Sun}},
  \bibinfo{journal}{Phys. Rev. B} \textbf{\bibinfo{volume}{75}},
  \bibinfo{pages}{033407} (\bibinfo{year}{2007}).

\bibitem[{\citenamefont{Zhou and Mizel}(2006)}]{Zhou2006Dec}
\bibinfo{author}{\bibfnamefont{X.}~\bibnamefont{Zhou}} \bibnamefont{and}
  \bibinfo{author}{\bibfnamefont{A.}~\bibnamefont{Mizel}},
  \bibinfo{journal}{Phys. Rev. Lett.} \textbf{\bibinfo{volume}{97}},
  \bibinfo{pages}{267201} (\bibinfo{year}{2006}).

\bibitem[{\citenamefont{Rabl et~al.}(2004)\citenamefont{Rabl, Shnirman, and
  Zoller}}]{Rabl2004Nov}
\bibinfo{author}{\bibfnamefont{P.}~\bibnamefont{Rabl}},
  \bibinfo{author}{\bibfnamefont{A.}~\bibnamefont{Shnirman}}, \bibnamefont{and}
  \bibinfo{author}{\bibfnamefont{P.}~\bibnamefont{Zoller}},
  \bibinfo{journal}{Phys. Rev. B} \textbf{\bibinfo{volume}{70}},
  \bibinfo{pages}{205304} (\bibinfo{year}{2004}).

\bibitem[{\citenamefont{Ruskov et~al.}(2005)\citenamefont{Ruskov, Schwab, and
  Korotkov}}]{Ruskov2005Nov}
\bibinfo{author}{\bibfnamefont{R.}~\bibnamefont{Ruskov}},
  \bibinfo{author}{\bibfnamefont{K.}~\bibnamefont{Schwab}}, \bibnamefont{and}
  \bibinfo{author}{\bibfnamefont{A.~N.} \bibnamefont{Korotkov}},
  \bibinfo{journal}{Phys. Rev. B} \textbf{\bibinfo{volume}{71}},
  \bibinfo{pages}{235407} (\bibinfo{year}{2005}).

\bibitem[{\citenamefont{Mancini et~al.}(2002)\citenamefont{Mancini,
  Giovannetti, Vitali, and Tombesi}}]{Mancini2002}
\bibinfo{author}{\bibfnamefont{S.}~\bibnamefont{Mancini}},
  \bibinfo{author}{\bibfnamefont{V.}~\bibnamefont{Giovannetti}},
  \bibinfo{author}{\bibfnamefont{D.}~\bibnamefont{Vitali}}, \bibnamefont{and}
  \bibinfo{author}{\bibfnamefont{P.}~\bibnamefont{Tombesi}},
  \bibinfo{journal}{Phys. Rev. Lett.} \textbf{\bibinfo{volume}{88}},
  \bibinfo{pages}{120401} (\bibinfo{year}{2002}).

\bibitem[{\citenamefont{Pirandola et~al.}(2006)\citenamefont{Pirandola, Vitali,
  Tombesi, and Lloyd}}]{Pirandola2006}
\bibinfo{author}{\bibfnamefont{S.}~\bibnamefont{Pirandola}},
  \bibinfo{author}{\bibfnamefont{D.}~\bibnamefont{Vitali}},
  \bibinfo{author}{\bibfnamefont{P.}~\bibnamefont{Tombesi}}, \bibnamefont{and}
  \bibinfo{author}{\bibfnamefont{S.}~\bibnamefont{Lloyd}},
  \bibinfo{journal}{Phys. Rev. Lett.} \textbf{\bibinfo{volume}{97}},
  \bibinfo{pages}{150403} (\bibinfo{year}{2006}).

\bibitem[{\citenamefont{Eisert et~al.}(2004)\citenamefont{Eisert, Plenio, Bose,
  and Hartley}}]{Eisert2004}
\bibinfo{author}{\bibfnamefont{J.}~\bibnamefont{Eisert}},
  \bibinfo{author}{\bibfnamefont{M.~B.} \bibnamefont{Plenio}},
  \bibinfo{author}{\bibfnamefont{S.}~\bibnamefont{Bose}}, \bibnamefont{and}
  \bibinfo{author}{\bibfnamefont{J.}~\bibnamefont{Hartley}},
  \bibinfo{journal}{Phys. Rev. Lett.} \textbf{\bibinfo{volume}{93}},
  \bibinfo{pages}{190402} (\bibinfo{year}{2004}).

\bibitem[{\citenamefont{Hu and Nori}(1996{\natexlab{a}})}]{Hu1995}
\bibinfo{author}{\bibfnamefont{X.}~\bibnamefont{Hu}} \bibnamefont{and}
  \bibinfo{author}{\bibfnamefont{F.}~\bibnamefont{Nori}},
  \bibinfo{journal}{Phys. Rev. Lett.} \textbf{\bibinfo{volume}{76}},
  \bibinfo{pages}{2294} (\bibinfo{year}{1996}{\natexlab{a}}).

\bibitem[{\citenamefont{Hu and Nori}(1996{\natexlab{b}})}]{Hu1996}
\bibinfo{author}{\bibfnamefont{X.}~\bibnamefont{Hu}} \bibnamefont{and}
  \bibinfo{author}{\bibfnamefont{F.}~\bibnamefont{Nori}},
  \bibinfo{journal}{Phys. Rev. B} \textbf{\bibinfo{volume}{53}},
  \bibinfo{pages}{2419} (\bibinfo{year}{1996}{\natexlab{b}}).

\bibitem[{\citenamefont{Hu and Nori}(1997)}]{Hu1997}
\bibinfo{author}{\bibfnamefont{X.}~\bibnamefont{Hu}} \bibnamefont{and}
  \bibinfo{author}{\bibfnamefont{F.}~\bibnamefont{Nori}},
  \bibinfo{journal}{Phys. Rev. Lett.} \textbf{\bibinfo{volume}{79}},
  \bibinfo{pages}{4605} (\bibinfo{year}{1997}).

\bibitem[{\citenamefont{Hu and Nori}(1999)}]{Hu1999}
\bibinfo{author}{\bibfnamefont{X.}~\bibnamefont{Hu}} \bibnamefont{and}
  \bibinfo{author}{\bibfnamefont{F.}~\bibnamefont{Nori}},
  \bibinfo{journal}{Physica B} \textbf{\bibinfo{volume}{263}},
  \bibinfo{pages}{16} (\bibinfo{year}{1999}).

\bibitem[{\citenamefont{You and Nori}(2005)}]{You2005PT}
\bibinfo{author}{\bibfnamefont{J.~Q.} \bibnamefont{You}} \bibnamefont{and}
  \bibinfo{author}{\bibfnamefont{F.}~\bibnamefont{Nori}},
  \bibinfo{journal}{Physics Today} \textbf{\bibinfo{volume}{58}}~(11),
  \bibinfo{pages}{42} (\bibinfo{year}{2005}).

\bibitem[{\citenamefont{Wendin and Shumeiko}(2005)}]{Wendin2005}
\bibinfo{author}{\bibfnamefont{G.}~\bibnamefont{Wendin}} \bibnamefont{and}
  \bibinfo{author}{\bibfnamefont{V.}~\bibnamefont{Shumeiko}},
  \bibinfo{journal}{cond-mat/0508729}  (\bibinfo{year}{2005}).

\bibitem[{\citenamefont{Zavatta et~al.}(2004)\citenamefont{Zavatta, Viciani,
  and Bellini}}]{Zavatta2004}
\bibinfo{author}{\bibfnamefont{A.}~\bibnamefont{Zavatta}},
  \bibinfo{author}{\bibfnamefont{S.}~\bibnamefont{Viciani}}, \bibnamefont{and}
  \bibinfo{author}{\bibfnamefont{M.}~\bibnamefont{Bellini}},
  \bibinfo{journal}{Science} \textbf{\bibinfo{volume}{306}},
  \bibinfo{pages}{660} (\bibinfo{year}{2004}).

\bibitem[{\citenamefont{Day et~al.}(2003)\citenamefont{Day, LeDuc, Mazin,
  Vayonakis, and Zmuidzinas}}]{Day2003}
\bibinfo{author}{\bibfnamefont{P.~K.} \bibnamefont{Day}},
  \bibinfo{author}{\bibfnamefont{H.~G.} \bibnamefont{LeDuc}},
  \bibinfo{author}{\bibfnamefont{B.~A.} \bibnamefont{Mazin}},
  \bibinfo{author}{\bibfnamefont{A.}~\bibnamefont{Vayonakis}},
  \bibnamefont{and}
  \bibinfo{author}{\bibfnamefont{J.}~\bibnamefont{Zmuidzinas}},
  \bibinfo{journal}{Nature} \textbf{\bibinfo{volume}{425}},
  \bibinfo{pages}{817} (\bibinfo{year}{2003}).

\bibitem[{\citenamefont{Scully and Zubairy}(1997)}]{Scullybook1997}
\bibinfo{author}{\bibfnamefont{M.~O.} \bibnamefont{Scully}} \bibnamefont{and}
  \bibinfo{author}{\bibfnamefont{M.~S.} \bibnamefont{Zubairy}},
  \emph{\bibinfo{title}{Quantum Optics}} (\bibinfo{publisher}{Cambridge
  University Press, Cambridge}, \bibinfo{year}{1997}).

\bibitem[{\citenamefont{Gaidarzhy
  et~al.}(2005{\natexlab{c}})\citenamefont{Gaidarzhy, Zolfagharkhani, Badzey,
  and Mohanty}}]{Gaidarzhy2005June}
\bibinfo{author}{\bibfnamefont{A.}~\bibnamefont{Gaidarzhy}},
  \bibinfo{author}{\bibfnamefont{G.}~\bibnamefont{Zolfagharkhani}},
  \bibinfo{author}{\bibfnamefont{R.~L.} \bibnamefont{Badzey}},
  \bibnamefont{and} \bibinfo{author}{\bibfnamefont{P.}~\bibnamefont{Mohanty}},
  \bibinfo{journal}{App. Phys. Lett.} \textbf{\bibinfo{volume}{86}},
  \bibinfo{pages}{254103} (\bibinfo{year}{2005}{\natexlab{c}}).

\bibitem[{\citenamefont{Valenzuela et~al.}(2006)\citenamefont{Valenzuela,
  Oliver, Berns, Berggren, Levitov, and Orlando}}]{Valenzuela2006Dec}
\bibinfo{author}{\bibfnamefont{S.~O.} \bibnamefont{Valenzuela}},
  \bibinfo{author}{\bibfnamefont{W.~D.} \bibnamefont{Oliver}},
  \bibinfo{author}{\bibfnamefont{D.~M.} \bibnamefont{Berns}},
  \bibinfo{author}{\bibfnamefont{K.~K.} \bibnamefont{Berggren}},
  \bibinfo{author}{\bibfnamefont{L.~S.} \bibnamefont{Levitov}},
  \bibnamefont{and} \bibinfo{author}{\bibfnamefont{T.~P.}
  \bibnamefont{Orlando}}, \bibinfo{journal}{Science}
  \textbf{\bibinfo{volume}{314}}, \bibinfo{pages}{1589} (\bibinfo{year}{2006}).

\bibitem[{\citenamefont{LaHaye et~al.}(2004)\citenamefont{LaHaye, Buu,
  Camarota, and Schwab}}]{LaHaye2004}
\bibinfo{author}{\bibfnamefont{M.~D.} \bibnamefont{LaHaye}},
  \bibinfo{author}{\bibfnamefont{O.}~\bibnamefont{Buu}},
  \bibinfo{author}{\bibfnamefont{B.}~\bibnamefont{Camarota}}, \bibnamefont{and}
  \bibinfo{author}{\bibfnamefont{K.~C.} \bibnamefont{Schwab}},
  \bibinfo{journal}{Science} \textbf{\bibinfo{volume}{304}},
  \bibinfo{pages}{74} (\bibinfo{year}{2004}).

\bibitem{Munday2005}
\bibinfo{author}{\bibfnamefont{J.~N.} \bibnamefont{Munday}},
  \bibinfo{author}{\bibfnamefont{D.}~\bibnamefont{Iannuzzi}},
  \bibinfo{author}{\bibfnamefont{Y.}~\bibnamefont{Barash}},
  \bibnamefont{and}
  \bibinfo{author}{\bibfnamefont{F.}~\bibnamefont{Capasso}},
  \bibinfo{journal}{Phys. Rev. A.}
  \textbf{\bibinfo{volume}{71}}, \bibinfo{pages}{042102}
  (\bibinfo{year}{2005}).

\bibitem{Munday2006}
\bibinfo{author}{\bibfnamefont{J.~N.} \bibnamefont{Munday}},
  \bibinfo{author}{\bibfnamefont{D.}~\bibnamefont{Iannuzzi}},
  \bibnamefont{and}
  \bibinfo{author}{\bibfnamefont{F.}~\bibnamefont{Capasso}},
  \bibinfo{journal}{New J. of Phys.}
  \textbf{\bibinfo{volume}{8}}, \bibinfo{pages}{244}
  (\bibinfo{year}{2006}).

\bibitem{Capasso2007}
\bibinfo{author}{\bibfnamefont{F.} \bibnamefont{Capasso}},
  \bibinfo{author}{\bibfnamefont{J.~N.}~\bibnamefont{Munday}},
  \bibinfo{author}{\bibfnamefont{D.}~\bibnamefont{Iannuzzi}}, \bibnamefont{and}
  \bibinfo{author}{\bibfnamefont{H.~B.} \bibnamefont{Chan}},
  \bibinfo{journal}{IEEE J. of Selected Topics in Quantum Electronics}
  \textbf{\bibinfo{volume}{13}},
  \bibinfo{pages}{400} (\bibinfo{year}{2007}).

\bibitem[{\citenamefont{Cooper et~al.}(2004)\citenamefont{Cooper, Steffen,
  McDermott, Simmonds, Oh, Hite, Pappas, and Martinis}}]{Cooper2004}
\bibinfo{author}{\bibfnamefont{K.~B.} \bibnamefont{Cooper}},
  \bibinfo{author}{\bibfnamefont{M.}~\bibnamefont{Steffen}},
  \bibinfo{author}{\bibfnamefont{R.}~\bibnamefont{McDermott}},
  \bibinfo{author}{\bibfnamefont{R.~W.} \bibnamefont{Simmonds}},
  \bibinfo{author}{\bibfnamefont{S.}~\bibnamefont{Oh}},
  \bibinfo{author}{\bibfnamefont{D.~A.} \bibnamefont{Hite}},
  \bibinfo{author}{\bibfnamefont{D.~P.} \bibnamefont{Pappas}},
  \bibnamefont{and} \bibinfo{author}{\bibfnamefont{J.~M.}
  \bibnamefont{Martinis}}, \bibinfo{journal}{Phys. Rev. Lett.}
  \textbf{\bibinfo{volume}{93}}, \bibinfo{pages}{180401}
  (\bibinfo{year}{2004}).

\bibitem[{\citenamefont{Naik et~al.}(2006)\citenamefont{Naik, Buu, LaHaye,
  Armour, Clerk, Blencowe, and Schwab}}]{Naik2006}
\bibinfo{author}{\bibfnamefont{A.}~\bibnamefont{Naik}},
  \bibinfo{author}{\bibfnamefont{O.}~\bibnamefont{Buu}},
  \bibinfo{author}{\bibfnamefont{M.~D.} \bibnamefont{LaHaye}},
  \bibinfo{author}{\bibfnamefont{A.~D.} \bibnamefont{Armour}},
  \bibinfo{author}{\bibfnamefont{A.~A.} \bibnamefont{Clerk}},
  \bibinfo{author}{\bibfnamefont{M.~P.} \bibnamefont{Blencowe}},
  \bibnamefont{and} \bibinfo{author}{\bibfnamefont{K.~C.}
  \bibnamefont{Schwab}}, \bibinfo{journal}{Nature}
  \textbf{\bibinfo{volume}{443}}, \bibinfo{pages}{193} (\bibinfo{year}{2006}).

\bibitem[{\citenamefont{Schliesser et~al.}(2006)\citenamefont{Schliesser,
  Del'Haye, Nooshi, Vahala, and Kippenberg}}]{Schliesser2006}
\bibinfo{author}{\bibfnamefont{A.}~\bibnamefont{Schliesser}},
  \bibinfo{author}{\bibfnamefont{P.}~\bibnamefont{Del'Haye}},
  \bibinfo{author}{\bibfnamefont{N.}~\bibnamefont{Nooshi}},
  \bibinfo{author}{\bibfnamefont{K.~J.} \bibnamefont{Vahala}},
  \bibnamefont{and} \bibinfo{author}{\bibfnamefont{T.~J.}
  \bibnamefont{Kippenberg}}, \bibinfo{journal}{Phys. Rev. Lett.}
  \textbf{\bibinfo{volume}{97}}, \bibinfo{pages}{243905}
  (\bibinfo{year}{2006}).

\bibitem[{\citenamefont{Kleckner and Bouwmeester}(2006)}]{Kleckner2006}
\bibinfo{author}{\bibfnamefont{D.}~\bibnamefont{Kleckner}} \bibnamefont{and}
  \bibinfo{author}{\bibfnamefont{D.}~\bibnamefont{Bouwmeester}},
  \bibinfo{journal}{Nature (London)} \textbf{\bibinfo{volume}{444}},
  \bibinfo{pages}{75} (\bibinfo{year}{2006}).

\bibitem[{\citenamefont{Gigan et~al.}(2006)\citenamefont{Gigan, Bohm,
  Paternostro, Blaser, Langer, Hertzberg, Schwab, Bauerle, Aspelmeyer, and
  Zeilinger}}]{Gigan2006}
\bibinfo{author}{\bibfnamefont{S.}~\bibnamefont{Gigan}},
  \bibinfo{author}{\bibfnamefont{H.~R.} \bibnamefont{Bohm}},
  \bibinfo{author}{\bibfnamefont{M.}~\bibnamefont{Paternostro}},
  \bibinfo{author}{\bibfnamefont{F.}~\bibnamefont{Blaser}},
  \bibinfo{author}{\bibfnamefont{G.}~\bibnamefont{Langer}},
  \bibinfo{author}{\bibfnamefont{J.~B.} \bibnamefont{Hertzberg}},
  \bibinfo{author}{\bibfnamefont{K.~C.} \bibnamefont{Schwab}},
  \bibinfo{author}{\bibfnamefont{D.}~\bibnamefont{Bauerle}},
  \bibinfo{author}{\bibfnamefont{M.}~\bibnamefont{Aspelmeyer}},
  \bibnamefont{and}
  \bibinfo{author}{\bibfnamefont{A.}~\bibnamefont{Zeilinger}},
  \bibinfo{journal}{Nature (London)} \textbf{\bibinfo{volume}{444}},
  \bibinfo{pages}{67} (\bibinfo{year}{2006}).

\bibitem[{\citenamefont{Poggio et~al.}(2007)\citenamefont{Poggio, Degen, Mamin,
  and Rugar}}]{Poggio2007}
\bibinfo{author}{\bibfnamefont{M.}~\bibnamefont{Poggio}},
  \bibinfo{author}{\bibfnamefont{C.~L.} \bibnamefont{Degen}},
  \bibinfo{author}{\bibfnamefont{H.~J.} \bibnamefont{Mamin}}, \bibnamefont{and}
  \bibinfo{author}{\bibfnamefont{D.}~\bibnamefont{Rugar}},
  \bibinfo{journal}{cond-mat/0702446}  (\bibinfo{year}{2007}).

\bibitem{Irish2003}
\bibinfo{author}{\bibfnamefont{E. K.}~\bibnamefont{Irish}} \bibnamefont{and}
  \bibinfo{author}{\bibfnamefont{K. C.}~\bibnamefont{Schwab}},
  \bibinfo{journal}{Phys. Rev. B} \textbf{\bibinfo{volume}{68}},
  \bibinfo{pages}{155311} (\bibinfo{year}{2003}).

\bibitem[{\citenamefont{Zhang et~al.}(2003)\citenamefont{Zhang, Peng, and
  Braunstein}}]{Zhang2003}
\bibinfo{author}{\bibfnamefont{J.}~\bibnamefont{Zhang}},
  \bibinfo{author}{\bibfnamefont{K.}~\bibnamefont{Peng}}, \bibnamefont{and}
  \bibinfo{author}{\bibfnamefont{S.~L.} \bibnamefont{Braunstein}},
  \bibinfo{journal}{Phys. Rev. A} \textbf{\bibinfo{volume}{68}},
  \bibinfo{pages}{013808} (\bibinfo{year}{2003}).

\end{thebibliography}

\end{document}